\documentclass[%
 reprint,
 amsmath,amssymb,
 aps,
]{revtex4-2}
\usepackage{latexsym,amsmath,amssymb,amscd,amsfonts,amsthm, xcolor}

\usepackage{caption,float}

\newcommand{\bs}{\left\{}
\newcommand{\es}{\right.}
\newcommand{\ba}{\begin{array}}
\newcommand{\ea}{\end{array}}
\newcommand{\be}{\begin{equation}}
\newcommand{\ee}{\end{equation}}
\newcommand{\R}{\mathbb{R}}

\newcommand{\myhash}{\raisebox{\depth}{\#}}

\usepackage{graphicx}
\usepackage{dcolumn}
\usepackage{bm}

\begin{document}
\title{Analytic expression of the DOS for a new model of 1d-potential and its random perturbation}
\author{Hakim Boumaza}
\affiliation{LAGA, Institut Galil\'ee, Universit\'e Sorbonne Paris Nord, 99 avenue J.B. Cl\'ement, F-93430 Villetaneuse}
\author{Olivier Lafitte}
\affiliation{IRL CRM-CNRS, Universit\'e de Montr\'eal, Pavillon Aisenstadt, 2920 Chemin de la tour, Montr\'eal (Qu\'ebec, Canada) H3T 1J4}


\date{}

\maketitle



\section{Introduction}

Density of states (DOS) plays an important role in various physical systems which have large number of particles confined in a given volume. Explicit or numerical calculations of the DOS are in general used to compare models with experiments, but very few explicit models exist. Indeed, apart from models where the dispersion relation is spherically symmetric and monotonically rising, it is complicated to invert it to express the quantum number $k$ as a function of the energy which would allow to obtain the DOS by a derivation of the $k$-space volume.

Among all the references on this subject, mention reference \cite{numerics-XPS} which experimentally observe this valence band using X-ray photoelectron spectroscopy and restore the DOS, which presents multiple peaks. The spinless fermion model \cite{numerics-fermion} compares the Luttinger liquids and the Fermi liquids. The density of states $D(\epsilon)$, where $\epsilon$ is the (low) energy level considered) is then computed numerically, using a 400-site chain, for different values of an interaction parameter in some  interval $[-2t,2t]$.  They obtain numerically a behavior of the form $D(\epsilon)\simeq \vert \epsilon-\epsilon_f\vert^{\alpha}$, where $\alpha$ should be measured, knowing that the non-interacting chain has a DOS equal to $\frac{1}{\pi}\frac{1}{\sqrt{4t^2-\epsilon^2}}1_{\vert \epsilon \vert<2t}$. This makes this study close to the type of behavior our paper addresses.\\

\begin{figure}[ht]
\begin{center}
\includegraphics[width=70mm]{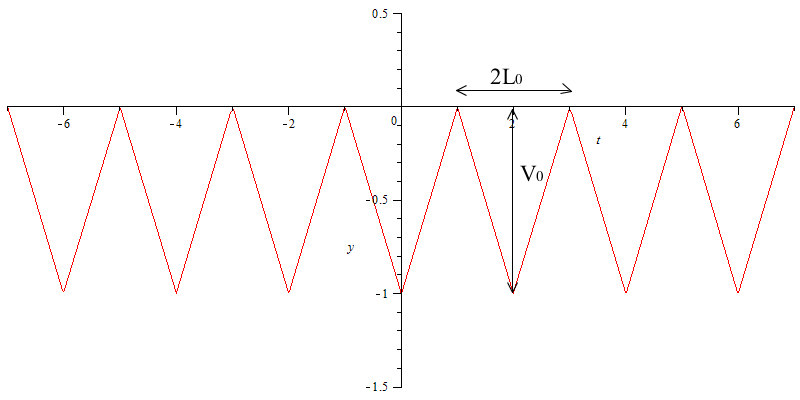} 
\caption{The considered periodic potential $\mathbf{V}$}\label{fig1}
\end{center}
\end{figure}

Our analysis focuses on comparisons between the spectrum of a one-dimensional Schr\"odinger operator for a particular periodic potential $\mathbf{V}$ (infinite in space) and for its restriction to a finite number of sites (see Figure \ref{fig1}). We deduce from this finite, but large, number of sites, the Integrated Density of States (IDS) associated to the Hamiltonian operator whose derivate is the DOS. The exact formula for the IDS is given in the Appendix. The expression of the DOS is analytical and, as the names suggest, is the derivative of the IDS with respect to the energy level $E$ is the DOS. All our calculations are done on the particular potential $\mathbf{V}$, which is a new case for which one has an analytical expression of the DOS. It is a continuous, periodic potential of period $2L_0$, of range $[-V_0,0]$, piecewise affine (which spectrum is, classically a band spectrum). For computing the DOS, we study the energy levels of the Schr\"odinger operator with the potential $\mathbf{V_{2N+1}}:=\mathbf{V}1_{[-(2N+1)L_0, (2N+1)L_0]}$ where $1_{[-(2N+1)L_0, (2N+1)L_0]}$ is the characteristic function of the interval $[-(2N+1)L_0, (2N+1)L_0]$ equal to $1$ on it and $0$ elsewhere. We state all the results we obtain in a nutshell:

\noindent 1. All eigenvalues of the finite range potential $\mathbf{V_{2N+1}}$ belong to the bands of the periodic potential (see \cite{hakim-olivier2}).

\noindent 2. The number of eigenvalues smaller than a given energy level $E$ is equivalent, when the number $N$ of atoms goes to infinity, to $2L_0(2N+1)\mathrm{IDS}(E)$ (see \eqref{IDS}, which is the result of a calculus based on certain special functions).

\noindent 3. The derivative of the IDS is singular at the edges $E_b$ of the bands with a singularity $|E-E_b|^{-\frac12}$. It was obtained numerically in \cite{numerics-fermion} for the Fermion 1d model,

\section{DOS representation and properties}
\subsection{The analytic solution}
Introduce the dimensionless constant
$$\kappa=\left(\frac{2mL_0^2\mathbf{V_0}}{\hbar^2}\right)^{\frac13}$$
This constant is characteristic of the lattice.

\begin{figure}[ht]
\begin{center}
\includegraphics[width=80mm]{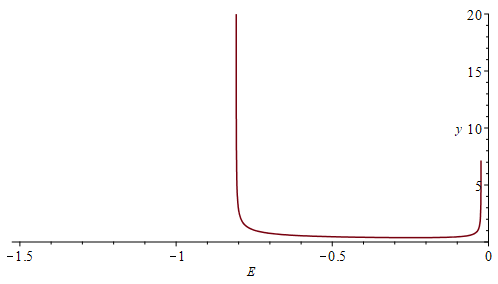} 
\caption{The DOS in the Hydrogen case ($\mathbf{V_0}=13,6$eV, $L_0=2\mathring{A}$ and $\kappa=1.526$)}\label{fig4}
\end{center}
\end{figure}
\begin{figure}[ht]

\begin{center}
\includegraphics[width=80mm]{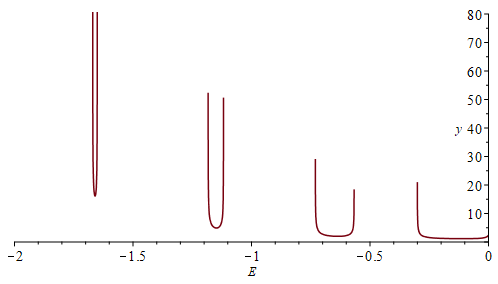} 
\caption{The DOS in the Carbon case ($\mathbf{V_0}=489.99$eV, $L_0=3.08\mathring{A}$ and $\kappa=10.682$)}\label{fig5}
\end{center}
\end{figure}

Recall that the spectrum of the operator $H=-\frac{\hbar^2}{2m}\frac{d^2}{dx^2}+\mathbf{V}$ (of domain  $H^2(\R)$) is a \textbf{band spectrum} $\cup_{j=0}^{+\infty}[E^j_{min}, E^j_{max}]$, containing only elements of the continuous spectrum, each of which being associated with a pseudo-eigenfunction (a non necessarily square summable function solution of $H\Psi_E=E\Psi_E$), whereas the spectrum of the operator $H_{2N+1}= -\frac{\hbar^2}{2m}\frac{d^2}{dx^2}+\mathbf{V_{2N+1}}$, of domain $D(H_{2N+1})=\{\psi\in H^2(\R), \psi(-(2N+1)L_0)=\psi((2N+1)L_0)\}$ is a pure point spectrum. By convention, we call ``level of energies of $\mathbf{V_{2N+1}}$'' the eigenvalues of $H_{2N+1}$ and ``spectral bands or gaps of $\mathbf{V}$'' the spectral bands and spectral gaps of $H$.

It is proven in \cite{hakim-olivier2} that there is no level of energy of $\mathbf{V_{2N+1}}$ in the spectral gaps of $\mathbf{V}$ and there are exactly $2N+2$ level of energies of $\mathbf{V_{2N+1}}$ in each spectral band of $\mathbf{V}$. 

A normalized counting function of the eigenvalues of $H_{2N+1}$ reads
$$I_N(E)=\frac{1}{2L_0(2N+1)}\myhash\{\lambda\leq E, \lambda\in \sigma(H_{2N+1})\}.$$
The {\bf Integrated Density of States} (IDS for short) is the limit of $I_N(E)$ as $N\rightarrow +\infty$. Note that in the normalization factor $2L_0(2N+1)$, $2N+1$ is the number of atoms in the finite lattice.

The IDS is a natural object defined through Birkhoff's ergodic theorem and it is easier to compute and study than its derivative which is  defined as the mathematical DOS, whenever possible.

 Introduce $\mathsf{U}(x)=\pi(Bi'(-x-\kappa\tfrac{E}{\mathbf{V_0}})Ai(x)-Ai'(-x-\kappa\tfrac{E}{\mathbf{V_0}})Bi(x))$ and $\mathsf{V}(x)=\pi(Ai(-x-\kappa\tfrac{E}{\mathbf{V_0}})Bi(x)-Bi(-x-\kappa\tfrac{E}{\mathbf{V_0}})Ai(x)),$ as well as the function $\Phi$, defined on $\sigma(H)$ through
$$\Phi(E)=2\arctan{\sqrt{-\frac{\mathsf{U}'\mathsf{V}}{\mathsf{U}\mathsf{V}'}}}\left(-\tfrac{\kappa E}{\mathbf{V_0}}\right),$$
were, incidentally, we notice that 
$$\sigma(H)=\{ E, (\mathsf{U}\mathsf{U}'\mathsf{V}\mathsf{V}')\left(\tfrac{\kappa E}{\mathbf{V_0}}\right)\leq 0\},$$
the edges of the bands $E^i_{min}, E^i_{max}$ being the roots of $\mathsf{U}\mathsf{U}'\mathsf{V}\mathsf{V}'$.
The derivative of the IDS given by \eqref{IDS} yields, for all real numbers $E$ and all $\kappa \geq \kappa_0$:
\be\label{DOS}
\hskip-2mm \mathrm{DOS}(E)=(-1)^{p(\frac{\kappa E}{\mathbf{V_0}}) }\frac{1}{2\pi}\frac{\kappa}{\mathbf{V_0}}\Phi'(\tfrac{\kappa E}{\mathbf{V_0}})1_{[E^{p(\frac{\kappa E}{\mathbf{V_0}})}_{min}, E^{p(\frac{\kappa E}{\mathbf{V_0}})}_{max}]}\ee
where $p(\frac{\kappa E}{\mathbf{V_0}})=[\frac{4}{3\pi}\kappa^{\frac32}(1+\frac{E}{\mathbf{V_0}})^{\frac32}]$. Note that $p(\frac{\kappa E}{\mathbf{V_0}})=1$ corresponds to the conduction band.

Recall that for generic Schr\"odinger operators with periodic potentials, the density of states behave, at the bottom of the spectrum, as $C(E-E_{min}^0)^{-\frac12}$ with $C>0$ a constant. In general nothing is known about the other bands. For a mathematical proof, see \cite[Theorem 2.1]{KS87} where the assumption of non-degeneracy of the spectral edges needed to apply \cite[Proposition 1.1]{K99} is proven. See also \cite{BN} for the non-degeneracy of the bottom of the continuous spectrum in the dimension $1$ case. Our potential $\mathbf{V}$ yields an example of periodic Schr\"odinger operator for which the IDS behaves as $K(E_b)|E-E_b|^{\frac12}$ at all edges $E_b$ of the spectral bands (see (\ref{eq_Ki})). By derivation of our explicit expression \eqref{IDS}, we recover that  the DOS has asymptotics $|E-E_b|^{-\frac12}$  at the edges of all spectral bands. 

From the expression \eqref{DOS} or from the Figure \ref{fig3}, one notice that when $N$ gets large, the eigenvalues of $H_{2N+1}$ accumulate in the spectral bands. They accumulate more at the edges of the bands than in the middle of the bands. One also notice that the bands gets thinner when they approach the bottom of the spectrum and also when $\kappa$, our dimensionless parameter, gets larger.

The formula \eqref{DOS} allows us to add a simple model, analytically tractable, which does not need numerical computations, (apart from plotting a classical function deduced from the Airy functions), in the list of 1d models one could hope to compare to the experimental data.

\subsection{Adding a random perturbation}
Another related situation worth mentioning is the behavior of the IDS in the presence of a random perturbation. In this case, the behavior of the IDS  changes drastically compared to the deterministic periodic case. In 1963, Lifshitz  had conjectured that, for a continuous random Schr\"odinger operator acting on $L^2(\R)$, there exist $c_{1},c_{2}>0$ such that its IDS satisfies the asymptotic $c_1\exp({-c_2(E-E_{0})^{-\frac{1}{2}}})$ as $E$ tends to $E_0$, where $E_0$ is the bottom of the spectrum of the considered Schr\"odinger operator (see \cite{L63}). This asymptotic behavior is known as Lifshitz tails and the exponent $-1/2$ is called the Lifshitz exponent of the operator.

Note that the IDS itself is defined through a thermodynamical limit which existence is proven through Birkhoff's ergodic theorem. Hence the IDS is almost-surely a deterministic quantity and it is not surprising that its asymptotic behavior at the bottom of the spectrum does not depend on the random parameters defining the random model.

The Lifschitz tails behavior has been proven first for the Poisson model (see \cite{DV75}) then for the Anderson model (see \cite{K99} for a general result about internal Lifschitz tails) and for the random displacement model (see \cite{KLNS}). The Lifshitz tails behavior means that the spectrum is very ``thin'' near $E_0$ and is a clue of the presence of the phenomenon of Anderson localization \cite{AW15}. Actually, the random models for which the Lifschitz tails behavior has been rigorously proven present Anderson localization in the sense that the almost-sure spectrum of these models is pure point and all eigenfuctions associated to eigenvalues in this pure point spectrum are exponentially decaying at infinity. And indeed, a pure point spectrum corresponds to the idea of a ``thin'' spectrum, by comparison to an absolutely continuous spectrum which contains full intervals as in the periodic case.

The random case is, from a mathematical point of view, better understood in the framework of the IDS than in the one of the DOS. If there is a vast literature about the IDS for random Schr\"odinger operators, very few is known for the DOS of the same models.

\section{Sketch of the proof}
The operator $H$ is a self-adjoint operator, of non compact resolvent (because the potential is not compactly supported). Hence it has  no pure point spectrum (see \cite{hakim-olivier1} where the results of Reed-Simon are recalled), and the absolutely continuous spectrum is composed of bands. Recall first that it has been proven in \cite{hakim-olivier1} that the bands of $H$ (segments of the purely absolutely continuous spectrum) are the $[E^i_{min}, E^i_{max}]$, where $E^{2i}_{min}, E^{2i+1}_{max}...$ are the eigenvalues of $-\frac{\hbar^2}{2m}\frac{d^2}{dx^2}+\mathbf{V}(x)$ on $H^1([-L_0, L_0])$ supplemented with the boundary conditions $\psi(L_0)=\psi(-L_0)$ and $\psi'(L_0)=\psi'(-L_0)$ and  $E^{2i}_{max}, E^{2i+1}_{min}...$, are the eigenvalues of $-\frac{\hbar^2}{2m}\frac{d^2}{dx^2}+\mathbf{V}(x)$ on $H^1([-L_0, L_0])$ supplemented with the boundary conditions $\psi(L_0)=-\psi(-L_0)$ and $\psi'(L_0)=-\psi'(-L_0)$.

The proofs are based upon the construction of a solution $\Psi_E$ of the eigenvalues equation $H\Psi_E = E \Psi_E$ usign transmission conditions between each period. We use there the uniqueness (up to a constant) of the solution outside the symmetric domain $[-(2N+1)L_0,(2N+1)L_0]$ and connect this solution to the inside of it.

Introduce the pair $( U, V)$ of fundamental solutions of the Airy equation. The functions $U(\lambda (\mathbf{V}(x)-E))$ and $\mathrm{sign}(\mathbf{V}(x)-E-[\mathbf{V}(x)-E])V(\lambda (\mathbf{V}(x)-E))$ are solution of $H$ outside the points $nL_0$, and construct a Bloch mode of $H$ if and only if an equality on $E$ using $U, V, U', V'$ is fulfilled (with four different equalities, see \cite{hakim-olivier1} for details). Extremely precise study of the behavior of the solutions of the Airy equations allow us to see the edges of the bands as all these ordered (interlaced) solutions of these equations.

It is then easy to define all eigenfuctions of $H_{2N+1}$ as being equal to $A_nU(\mathbf{V_{2N+1}}(x))+B_nV(\mathbf{V_{2N+1}}(x))$ for all $x\in [(2n-1)L_0, (2n+1)L_0]$. The regularity of the eigenfunction implies transfer conditions on $\left(\begin{smallmatrix}A_n\\ B_n \end{smallmatrix}\right)$, of the form $\left(\begin{smallmatrix} A_{n+1}\\ B_{n+1}\end{smallmatrix}\right)=T_{\frac{\kappa E}{\mathbf{V_0}}}\left(\begin{smallmatrix}A_n \\ B_n \end{smallmatrix}\right),$
supplemented with the conditions expressing the fact that the eigenfunction is a solution of the ODE $-\frac{\hbar^2}{2m}\frac{d^2}{dx^2}\psi=E\psi$ decaying at $\pm \infty$ outside the interval $[-(2N+1)L_0, (2N+1)L_0]$.
$$\left(\begin{smallmatrix} A_{-N}\\ B_{-N}\end{smallmatrix} \right)\propto\left(\begin{smallmatrix} 1\\
-\kappa^{\frac32}\sqrt{-\tfrac{E}{\mathbf{V_0}}}\end{smallmatrix} \right), \left(\begin{smallmatrix} A_{N}\\ B_{N}\end{smallmatrix} \right)\propto\left(\begin{smallmatrix} 1\\
\kappa^{\frac32}\sqrt{-\tfrac{E}{\mathbf{V_0}}}\end{smallmatrix} \right).$$
The eigenvalues $E$ are thus solutions of 
\begin{equation}\label{equa}T^{2N}\left(\begin{smallmatrix} 1 \\
\kappa^{\frac32}\sqrt{-\frac{E}{\mathbf{V_0}}}\end{smallmatrix}\right).\left(\begin{smallmatrix}
\kappa^{\frac32}\sqrt{-\frac{E}{\mathbf{V_0}}}\\
1\end{smallmatrix}\right)=0.
\end{equation}
This equation is explicitly written using again $U, V, U', V'$. These expressions allow us to study precisely the roots of the equation (\ref{equa}). The distribution of these roots then yields the IDS (the complete description of this method is done in \cite{hakim-olivier2}).

\section{Appendix: mathematical calculations}

We are able to describe the DOS and the IDS in the conduction band and in the valence band. The potential that we are able to treat is $\mathbf{V}$, the $2L_0$ periodic function which restriction to $[-L_0,L_0]$ is $-\mathbf{V_0}+\frac{\mathbf{V_0}}{L_0}\vert x\vert$.

For $E$ a real number, IDS$(E)$ is given by
\be\label{IDS}\frac12p(\frac{\kappa E}{\mathbf{V_0}})+\bs\ba{ll}\frac{1}{2\pi}\Phi(\frac{\kappa E}{\mathbf{V_0}})1_{[E^{p(\frac{\kappa E}{\mathbf{V_0}})}_{min}, E^{p(\frac{\kappa E}{\mathbf{V_0}})}_{max}]},& p(\frac{\kappa E}{\mathbf{V_0}}) \mbox{ even } \\
 (\frac12-\frac{1}{2\pi}\Phi(\frac{\kappa E}{\mathbf{V_0}}))1_{[E^{p(\frac{\kappa E}{\mathbf{V_0}})}_{min}, E^{p(\frac{\kappa E}{\mathbf{V_0}})}_{max}]},& p(\frac{\kappa E}{\mathbf{V_0}}) \mbox{ odd },\ea\es\ee
where $p(\frac{\kappa E}{\mathbf{V_0}})=[\frac{4}{3\pi}\kappa^{\frac32}(1+\frac{E}{\mathbf{V_0}})^{\frac32}]$ for $\kappa\geq \kappa_0$. 

Here, $-\kappa_0$ is the largest zero of $V$ introduced above ($\kappa_0\simeq 1.515$). This is proven in \cite{hakim-olivier2}. The presence of the integer multiple of $\tfrac{1}{2}=\tfrac{\pi}{2\pi}$ is also a consequence of the gap-labelling theorem (see \cite{J86}). For a simple explanation, see \cite[Page 390]{CL90}.

In the case $E$ not close to zero, one proves that ($i\leq p$)
\begin{equation}\label{eq_Ki}
\frac{\mathrm{IDS}(E)}{\sqrt{E-E^i_{min}}}=K^i_{min}(E), \frac{\mathrm{IDS}(E)}{\sqrt{E^i_{max}-E}}=K^i_{max}(E) 
\end{equation}
where the functions $K^i_{min}(E), K^i_{min}(E)$ are smooth, respectively,  in the neighborhood of $E^i_{min}, E^i_{max}$. One has respectively in the neighborhood of $E^i_{min}, E^i_{max} $, the existence of functions $R^i_{min}$, $R^i_{max}$ smooth  such that 
\begin{align*}
\mathrm{DOS}(E) & = (E-E^i_{min})^{-\frac12}R^i_{min}(E),\\
\mathrm{DOS}(E) & = (E^i_{max}-E)^{-\frac12}R^i_{max}(E).
\end{align*}

We also obtain the following exact expression equivalent to (\ref{IDS}), where $\tau_0=\frac{Ai}{Bi}(-\frac{cE}{\mathbf{V_0}}), \tau_1=\frac{Ai'}{Bi'}(-\frac{cE}{\mathbf{V_0}})$, $P$ and $Q$ are functions defined through the fractional Bessel functions (see \cite{O, hakim-olivier1}), $\zeta=\frac23(\kappa+E)^{\frac32}$, and $a,b,c,d$ are constructed with $P,Q, \cos \zeta, \sin \zeta$
$$\Phi(E)=\sqrt{\frac{a+\tau_0b-\tau_1c-\tau_0\tau_1d}{a+\tau_0c-\tau_1b-\tau_0\tau_1d}}\simeq \sqrt{ \frac{\tau_0-\frac{1}{\tau_1}-2\cos \zeta}{\tau_1-\frac{1}{\tau_0}+2\cos \zeta}}.$$

\begin{figure}[h]
\begin{center}
\includegraphics[width=70mm]{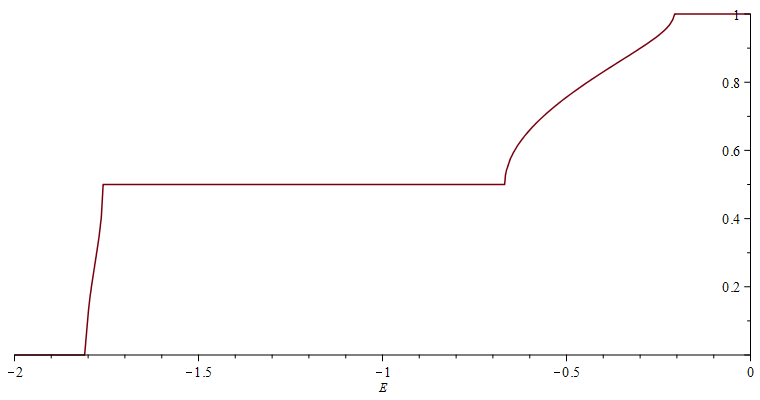} 
\end{center}
\caption{Integrated density of states for $\kappa=2.8$}\label{fig2}
\end{figure}

\begin{figure}[h]
\begin{center}
\includegraphics[width=70mm]{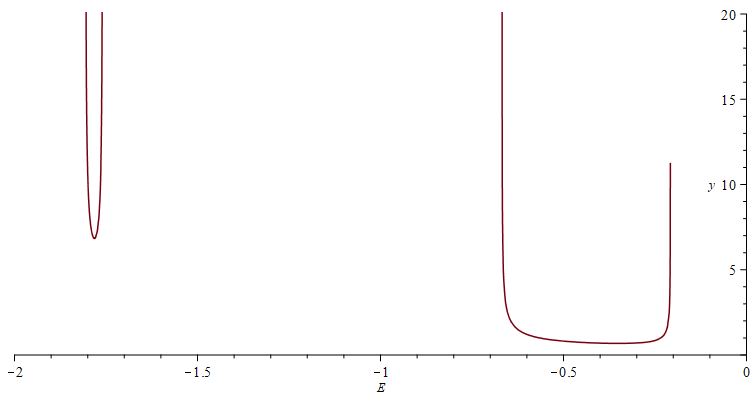} 
\caption{The DOS associated to the potential V for $\kappa=2.8$}\label{fig3}
\end{center}
\end{figure}

The graph for the IDS is given at Figure \ref{fig2}. This is the graph of a function, given by (\ref{IDS}). The graph of the DOS is given at Figure \ref{fig3}.

\begin{thebibliography}{99}

\bibitem{AW15} M. Aizenman, S. Warzel: Random operators. Disorder effects on quantum spectra and dynamics, Vol. 168 (Providence, RI : American Mathematical Society (AMS), 2015)

\bibitem{hakim-olivier1} H. Boumaza, O. Lafitte: The band spectrum of the periodic Airy-Schr\"odinger operator on the real line, {\em J. Differential Equations 264(1), 455-505 (2018), DOI:10.1016/j.jde.2017.09.013}
\bibitem{hakim-olivier2} H. Boumaza, O. Lafitte: Integrated density of states: from the finite range to the periodic Airy-Schr\"odinger operator, {\em J. Math. Phys. 62, 043503 (2021), https://doi.org/10.1063/5.0015181}
\bibitem{BN} H. Boumaza, H. Najar, Lifshitz tails for continuous matrix-valued Anderson models, {\em J. Stat. Phys. 160, No. 2, 371-396 (2015).} 
\bibitem{CL90} R. Carmona and J. Lacroix: Spectral Theory of Random Schr\"odinger Operators, {\em Probability and Its Applications, 
Birkh\"auser, Boston, (1990)}
\bibitem{DV75} M. Donsker, S. Varadhan: Asymptotic for the Wiener sausage, {\em Commun. Pure Appl. Math. 28, 525-565 (1975).}
\bibitem{numerics-fermion} E Jeckelmann: 
Local density of states of the one-dimensional spinless fermion model {\em J. Phys.: Condens. Matter 25 (2013) 014002}
\bibitem{J86} R. A. Johnson: Exponential dichotomy, rotation number, and linear differential operators with bounded coefficients. {\em J. Differential Equations, 61(1):54-78, (1986).}
\bibitem{KS87} W. Kirsch, B. Simon: Comparison theorems for the gap of Schr\"{o}dinger operators,{\em J. Funct. Anal. \textbf{75}(2), 396\textendash 410 (1987).}
\bibitem{K99} F. Klopp : Internal Lifshitz tails for random perturbations of periodic Schr\"{o}dinger operators, {\em Duke Math. J. \textbf{98}(2), 335\textendash 396 (1999).}
\bibitem{KLNS} F. Klopp, M. Loss, S. Nakamura, G. Stolz: Localization for the random displacement model, {\em Duke Math. J. 161, No. 4, 587-621 (2012).} 
\bibitem{numerics-XPS} Krasavin et al: Restoration of valence density of states from XPS spectra {\em IOP Conf. Series: Journal of Physics: Conf. Series 1923364(526071879)012014}
\bibitem{O} F. W. J. Olver, Asymptotics and special functions, Computer Science and Scientific Computing, Academic Press, San Diego, (1974).
\bibitem{L63} I. Lifshitz: Structure of the energy spectrum of impurity bands in disordered solid solutions, {\em Soviet Phy. JETP \textbf{17}, 1159-1170 (1963)}
\bibitem{band-structure} Masoud Seifikar et al: Self-consistent Green's function method for dilute nitride conduction band structure {\em 2014 J. Phys.: Condens. Matter 26 365502}
\end{thebibliography}
\end{document}